\pdfoutput=1
\pdfmapfile{+classico.map} 
\newif\ifafour
\afourtrue 

\documentclass[\ifafour a4paper,12pt,\else a5paper,10pt,\fi
onecolumn,oneside,article,
british%
]{memoir}
\newif\ifnotnotes
\notnotestrue 

\newcommand*{\oggi}{12 June 2017}
\newcommand*{\pdftitle}{Geometry of maximum-entropy proofs: stationary points, convexity, Legendre transforms, exponential families}
\newcommand*{\propertitle}{Geometry of maximum-entropy proofs:\\{\large
    stationary points, convexity,\\ Legendre transforms, exponential families\par}}
\newcommand*{\headtitle}{Geometry of maximum-entropy proofs}
\newcommand*{\pdfauthor}{P.G.L.  Porta Mana}
\newcommand*{\headauthor}{\ifnotnotes Porta Mana%
\else\autanet\ Luca\fi}
\newcommand*{\reporthead}{}

\usepackage[T1]{fontenc} 
\input{glyphtounicode} \pdfgentounicode=1
\usepackage[utf8]{inputenx}

\usepackage{textcomp}
\usepackage{empheq}
\usepackage{amssymb}
\usepackage{amsxtra}

\usepackage[main=british,french,italian,german,swedish,latin,esperanto]{babel}

\usepackage[autostyle=false,autopunct=false,english=british]{csquotes}
\setquotestyle{american}

\usepackage{amsthm}

\theoremstyle{remark}

\newtheoremstyle{innote}{\parsep}{\parsep}{\footnotesize}{}{}{}{0pt}{}
\theoremstyle{innote}

\usepackage[shortlabels,inline]{enumitem}
\SetEnumitemKey{para}{itemindent=\parindent,leftmargin=0pt,listparindent=\parindent,parsep=0pt,itemsep=\topsep}
\setlist[enumerate,2]{label=\alph*.}
\setlist[enumerate]{leftmargin=\parindent}
\setlist[itemize]{leftmargin=\parindent}
\setlist[description]{leftmargin=\parindent}

\usepackage[babel,theoremfont]{newpxtext}
\usepackage[bigdelims,nosymbolsc
]{newpxmath}
\useosf
\makeatletter
\def\re@DeclareMathSymbol#1#2#3#4{%
    \let#1=\undefined
    \DeclareMathSymbol{#1}{#2}{#3}{#4}}
\re@DeclareMathSymbol{\bigoplusop}{\mathop}{largesymbols}{"4C}
\re@DeclareMathSymbol{\bigotimesop}{\mathop}{largesymbols}{"4E}
\re@DeclareMathSymbol{\sumop}{\mathop}{largesymbols}{"50}
\re@DeclareMathSymbol{\prodop}{\mathop}{largesymbols}{"51}
\re@DeclareMathSymbol{\bigcupop}{\mathop}{largesymbols}{"53}
\re@DeclareMathSymbol{\bigcapop}{\mathop}{largesymbols}{"54}
\re@DeclareMathSymbol{\bigwedgeop}{\mathop}{largesymbols}{"56}
\re@DeclareMathSymbol{\bigveeop}{\mathop}{largesymbols}{"57}
\re@DeclareMathSymbol{\bigtimesop}{\mathop}{largesymbolsPXA}{"10}
\makeatother

\DeclareFontFamily{U}{egreek}{\skewchar\font'177}%
\DeclareFontShape{U}{egreek}{m}{n}{<-6>s*[1]eurm5 <6-8>s*[1]eurm7 <8->s*[1]eurm10}{}%
\DeclareFontShape{U}{egreek}{m}{it}{<->s*[1]eurmo10}{}%
\DeclareFontShape{U}{egreek}{b}{n}{<-6>s*[1]eurb5 <6-8>s*[1]eurb7 <8->s*[1]eurb10}{}%
\DeclareFontShape{U}{egreek}{b}{it}{<->s*[1]eurbo10}{}%
\DeclareSymbolFont{egreeki}{U}{egreek}{m}{it}%
\SetSymbolFont{egreeki}{bold}{U}{egreek}{b}{it}
\DeclareSymbolFont{egreekr}{U}{egreek}{m}{n}%
\SetSymbolFont{egreekr}{bold}{U}{egreek}{b}{n}
\DeclareFontFamily{U}{egreekx}{\skewchar\font'177}
\DeclareFontShape{U}{egreekx}{m}{n}{%
       <-7.5>s*[0.9]euex7%
    <7.5-8.5>s*[0.9]euex8%
    <8.5-9.5>s*[0.9]euex9%
    <9.5->s*[0.9]euex10%
}{}
\DeclareSymbolFont{egreekx}{U}{egreekx}{m}{n}
\DeclareMathSymbol{\sumop}{\mathop}{egreekx}{"50}
\DeclareMathSymbol{\prodop}{\mathop}{egreekx}{"51}
\DeclareMathSymbol{\coprodop}{\mathop}{egreekx}{"60}
\makeatletter
\def\sum{\DOTSI\sumop\slimits@}
\def\prod{\DOTSI\prodop\slimits@}
\def\coprod{\DOTSI\coprodop\slimits@}
\makeatother
 \DeclareMathSymbol{\partialup}{\mathalpha}{egreekr}{"40}
 \DeclareMathSymbol{\epsilon}{\mathalpha}{egreeki}{"0F}
 \DeclareMathSymbol{\varAlpha}{\mathalpha}{egreeki}{"41}
 \DeclareMathSymbol{\varGamma}{\mathalpha}{egreeki}{"00}
 \DeclareMathSymbol{\varEpsilon}{\mathalpha}{egreeki}{"45}
 \DeclareMathSymbol{\varSigma}{\mathalpha}{egreeki}{"06}
 \DeclareMathSymbol{\deltaup}{\mathalpha}{egreekr}{"0E}

\renewcommand\sfdefault{uop}
\DeclareMathAlphabet{\mathsf}  {T1}{\sfdefault}{m}{sl}
\SetMathAlphabet{\mathsf}{bold}{T1}{\sfdefault}{b}{sl}

\usepackage[scaled=0.84]{DejaVuSansMono}

\usepackage{mathdots}

\usepackage[usenames]{xcolor}
\definecolor{mybluishpurple}{RGB}{51,34,136}
\definecolor{myblue}{RGB}{136,204,238}
\definecolor{mybluishgreen}{RGB}{68,170,153}
\definecolor{mygreen}{RGB}{17,119,51}
\definecolor{mygreenishyellow}{RGB}{153,153,51}
\definecolor{myyellow}{RGB}{221,204,119}
\definecolor{myred}{RGB}{204,102,119}
\definecolor{mypurplishred}{RGB}{136,34,85}
\definecolor{myreddishpurple}{RGB}{170,68,153}

\usepackage{bm}
\usepackage{microtype}

\usepackage[backend=biber,mcite,
citestyle=authoryear-comp,bibstyle=pglpm-authoryear,autopunct=false,sorting=ny,sortcites=false,natbib=false,maxcitenames=1,maxbibnames=8,minbibnames=8,giveninits=true,uniquename=false,uniquelist=false,maxalphanames=1,block=space,hyperref=true,defernumbers=false,useprefix=true,sortupper=false,language=british,parentracker=false]{biblatex}
\DeclareSortingScheme{ny}{\sort{\field{sortname}\field{author}\field{editor}}\sort{\field{year}}}

\renewcommand*{\finalnamedelim}{, }
\DeclareDelimFormat{multicitedelim}{\addsemicolon\space}
\DeclareDelimFormat{compcitedelim}{\addsemicolon\space}
\DeclareDelimFormat{postnotedelim}{\space}
\renewcommand{\bibfont}{\footnotesize}
\newcommand*{\citep}{\parencites}
\newcommand*{\citey}{\parencites*}
\renewcommand*{\cites}{\parencites}
\providecommand{\href}[2]{#2}

\newcommand*{\amp}{\&}

\newcommand*{\subtitleproc}[1]{}

\usepackage{tensor}
\usepackage{graphicx}
\usepackage{wrapfig}

\usepackage{hyperref}
\usepackage[depth=4]{bookmark}
\hypersetup{colorlinks=true,bookmarksnumbered,pdfborder={0 0 0.25},citebordercolor={0.2 0.1333 0.5333},
citecolor=mybluishpurple,linkbordercolor={0.0667 0.4667 0.2},
linkcolor=mypurplishred,urlbordercolor={0.5333 0.1333 0.3333},
urlcolor=mygreen,breaklinks=true,pdftitle={\pdftitle},pdfauthor={\pdfauthor}}

\ifafour\setstocksize{297mm}{210mm}
\else\setstocksize{210mm}{5.5in}
\fi
\settrimmedsize{\stockheight}{\stockwidth}{*}
\setlxvchars[\normalfont] 
\setxlvchars[\normalfont]
\ifafour\settypeblocksize{*}{32pc}{1.618} 
\else\settypeblocksize{*}{26pc}{1.618}
\fi
\setulmargins{*}{*}{1}
\setlrmargins{*}{*}{*}
\setheadfoot{\onelineskip}{2.5\onelineskip}
\setheaderspaces{*}{2\onelineskip}{*}
\setmarginnotes{2ex}{10mm}{0pt}
\checkandfixthelayout[nearest]
\fixpdflayout

\newenvironment{acknowledgements}{\section*{Thanks}\addcontentsline{toc}{section}{Thanks}}{\par}
\counterwithout{section}{chapter}
\setsecnumformat{\upshape\csname the#1\endcsname\quad}
\setsecheadstyle{\bfseries\sffamily}
\setsubsecheadstyle{\bfseries\sffamily%
\raggedright}
\setbeforesecskip{3.25ex plus 1ex -.2ex}
\setaftersecskip{-1em}
\setaftersubsecskip{-1em}
\setsubsecindent{0pt}
\setparaheadstyle{\bfseries\sffamily%
\raggedright}
\newcommand{\addchap}[1]{\chapter*[#1]{#1}\addcontentsline{toc}{chapter}{#1}}
\newcommand{\addsec}[1]{\section*{#1}\addcontentsline{toc}{section}{#1}}

\defbibheading{bibliography}[\bibname]{\section*{{\normalsize #1}}\addcontentsline{toc}{section}{#1}\par
}

\copypagestyle{manaart}{plain}
\makeheadrule{manaart}{\headwidth}{0.5\normalrulethickness}
\makeoddhead{manaart}{%
{\footnotesize\sffamily%
\scshape\headauthor}}{}{{\footnotesize\sffamily%
\headtitle}}
\makeoddfoot{manaart}{}{\thepage}{}
\newcommand*\autanet{\includegraphics[height=\heightof{M}]{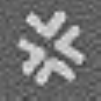}}
\definecolor{mygray}{gray}{0.333}
\ifafour
\else\fi
\makeoddfoot{plain}{}{\makebox[0pt]{\thepage}}{}
\makeoddhead{plain}{}{}{\footnotesize\reporthead}

\copypagestyle{manainitial}{plain}
\makeheadrule{manainitial}{\headwidth}{0.5\normalrulethickness}
\makeoddhead{manainitial}{%
\footnotesize\sffamily%
\scshape\headauthor}{}{\footnotesize\sffamily%
\headtitle}
\makeoddfoot{manaart}{}{\thepage}{}

\pretitle{\begin{center}\Large\sffamily%
\bfseries}
\posttitle{\bigskip\end{center}}

\makeatletter\newcommand*{\atf}{\includegraphics[
totalheight=\heightof{@}]{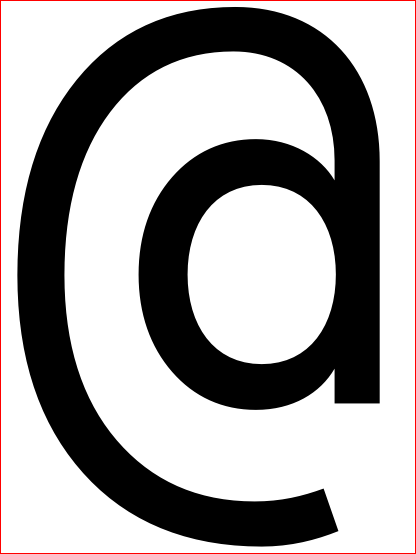}}\makeatother

\providecommand{\epost}[1]{\texttt{\footnotesize\textless#1\textgreater}}
\providecommand{\email}[2]{\href{mailto:#1ZZ@#2 ((remove ZZ))}{#1\protect\atf#2}}

\preauthor{\vspace{-0.5\baselineskip}\begin{center}
\normalsize\sffamily%
\lineskip  0.5em}
\postauthor{\par\end{center}}
\predate{\begin{center}\footnotesize}
\postdate{\end{center}\vspace{-\medskipamount}}
\usepackage[british]{datetime2}
\DTMnewdatestyle{mydate}%
{
}
\DTMsetdatestyle{mydate}

\setfloatadjustment{figure}{\footnotesize}
\captiondelim{\quad}
\captionnamefont{\footnotesize\sffamily%
}
\captiontitlefont{\footnotesize}
\firmlists*
\midsloppy

\title{\propertitle}
\author{\ifnotnotes%
P.G.L. Porta\,Mana%
\else Luca\fi
\quad
\epost{\email{mana}{kth.se}}%
}

\date{\oggi}

\newcommand*{\de}{\partialup}
\newcommand*{\RR}{\bm{\mathrm{R}}}
\newcommand*{\defd}{\coloneqq}
\renewcommand{\ge}{\geqslant}
\DeclarePairedDelimiter\clcl{[}{]}

\DeclarePairedDelimiter\set{\{}{\}}
\newcommand*{\p}{\mathrm{P}}
\newcommand*{\sect}{\S}
\newcommand*{\chap}{ch.}%
\newcommand*{\eqn}{eq.}%
\newcommand*{\eqns}{eqs}%
\newcommand*{\fig}{fig.}%
\newcommand*{\eg}{{e.g.}}

\newcommand*{\yp}{\bm{p}}
\newcommand*{\yq}{\bm{q}}
\newcommand*{\yll}{l}
\newcommand*{\yl}{\bm{\yll}}
\newcommand*{\ymm}{\varEpsilon}
\newcommand*{\yM}{\bm{\ymm}}
\newcommand*{\ycc}{\epsilon}
\newcommand*{\yc}{\bm{\ycc}}
\newcommand*{\ym}{u}
\newcommand*{\yu}{\bm{\varSigma}}
\newcommand*{\yF}{L}
\newcommand*{\yG}{\varGamma}

\newcommand*{\cd}{}

\firmlists
\begin{document}
\captiondelim{\quad}\captionnamefont{\footnotesize}\captiontitlefont{\footnotesize}
\selectlanguage{british}\frenchspacing

\maketitle

\selectlanguage{british}\frenchspacing


Most proofs of the traditional maximum-entropy formulae
\citep[\chap~5]{jaynes1963,sivia1996_r2006}
\begin{equation*}
  \begin{gathered}
    \yp=\frac{\exp(\yl\yM + \ln\yq)}{Z(\yl)}\qquad
    Z(\yl) = \yu\cd\exp(\yl\yM + \ln\yq)\qquad
    \frac{\de\ln Z}{\de\yl}=\yc,
  \end{gathered}
\end{equation*}
with or without $\yq$ terms, rely on the method of Lagrange multipliers,
and many of them also show that the constrained maximization of the Shannon
entropy is equivalent to the minimization of a \enquote{potential function}
via a Lagrange transform. Examples are Mead \amp\ Papanicolaou's
\citey[\sect~II]{meadetal1984} and Jaynes's
\citey[\sect~11.6]{jaynes1994_r2003} proofs.

This note is a geometric commentary on such proofs. Its purpose is to help
you visualize some of the geometric structures involved and to explain more
in detail why we end up \emph{minimizing} a potential function to
\emph{maximize} the entropy, and how Lagrange transforms emerge. A synopsis
of the main functions involved in the proof and of their very different
properties is given at the end, together with a brief discussion of
exponential families of probabilities, which appear in the proof.

I assume that you have some familiarity with the maximum-entropy method and
the standard proof of its formulae, as in the references above. The
geometric commentary is formulated in terms of the \emph{relative} Shannon
entropy \citep[\sect~4.c]{jaynes1963}{kullback1982_r2006,good1983b}, also
called negative discrimination information
\citep[\chap~3]{kullback1959_r1978}, with respect to a reference
distribution, because only this formulation is coordinate-independent in
the continuum limit \citep{hobsonetal1973,good1983b}.

\section{Notation} I hope you will indulge me in using vector-covector
notation, which produces compact multidimensional formulae. A covector maps
a vector into a scalar; it can be represented as a row matrix, and a vector
as a column matrix. Here Latin letters will represent vectors; Greek,
covectors. Juxtaposition of a vector and a covector always means their
contraction or row-column multiplication, irrespective of their order; \eg\
$\yc\yl=\yl\yc=\sum_n\ycc^n\yll_n$. I also use the convention that the
logarithm maps vectors into covectors elementwise, and vice versa for the
exponential; for example
$\ln\yp \defd (\ln p_1, \ln p_2, \dotsc)$ and is a covector.
Finally, \emph{convex} means $\cup$-shaped, and \emph{concave} means
$\cap$-shaped.

\section{Geometry of the proof} We have $K$ states $k\in\set{1,\dotsc, K}$
and $N$ observables $n\in\set{1,\dotsc,N}$. A measurement of the $n$th
observable when the state is $k$ gives the value $\ymm^{n k}$. These values
are grouped into the operator $\yM=(\varEpsilon^{n k})$ which maps vectors
to covectors. The distribution of probabilities $\yp=(p_k)$, a vector,
expresses our uncertainty about the actual state. The expectations for the
$N$ observables under our state of uncertainty are $\yM\yp$. The
$K$-dimensional covector $\yu=(1, \dotsc, 1)$ and allows us very suggestively
to write $\yu\cd\yp=\sum_k p_k$.

We want to choose a probability distribution for which the expectations are
constrained to values $\yc=(\ycc_n)$, represented by a covector. But the
constraints $\yM\cd\yp=\yc$ don't select a unique $\yp$ if $N<K-1$. We
therefore ask that that the distribution meet an additional requirement
that makes it unique: $\yp$ must maximize the relative Shannon entropy
$H(\yp) \defd -\yp\ln\yp+\yp\ln\yq$, under those constraints, with respect
to a reference distribution $\yq$. We denote this unique distribution
$\yp_{\yc}$, and the corresponding maximum value of the entropy
$S(\yc)\defd H(\yp_{\yc})$; it's called the \emph{Gibbs entropy}. The
Shannon and Gibbs entropies are different functions, of completely
different quantities.

To find the constrained maximum of the relative Shannon entropy we use the
method of Lagrange multipliers \citep[\chap~5]{boydetal2004_r2009}. I
warmly recommend Rockafellar's \citey{rockafellar1993} brilliant review of
the various meanings of Lagrange multipliers, which offers plenty of geometric
insights.

With scalar $\ym$ and vector $\yl = (\yll_n)$, define the function of
$(\yp,\ym,\yl)$ with parameter $\yc$
\begin{equation}
  \yF_{\yc}(\yp,\ym,\yl)\defd
  -\yp\ln\yp + \yp\ln\yq + \ym\,(\yu\yp-1) + \yl\,(\yM\yp-\yc),
  \label{eq:Lagrangian}
\end{equation}
usually called \emph{Lagrangian} \citep{fangetal1997,boydetal2004_r2009}.
It's defined on the $(K+N+1)$-dimensional manifold $
{\RR_{\ge0}}^{K}\times\RR^{N+1}$, our \enquote{base
  manifold}. Proofs of the maximum-entropy formulae with linear
constraints show that \emph{the Lagrangian $\yF_{\yc}(\yp,\ym,\yl )$ has a
  unique saddle point $(\yp_{\yc},\ym_{\yc},\yl_{\yc})$, and the
  saddle-point coordinate $\yp_{\yc}$ is the maximum-entropy solution}.

\begin{figure}[p!]
  {\centering\vspace{-1em}\includegraphics[width=\linewidth]{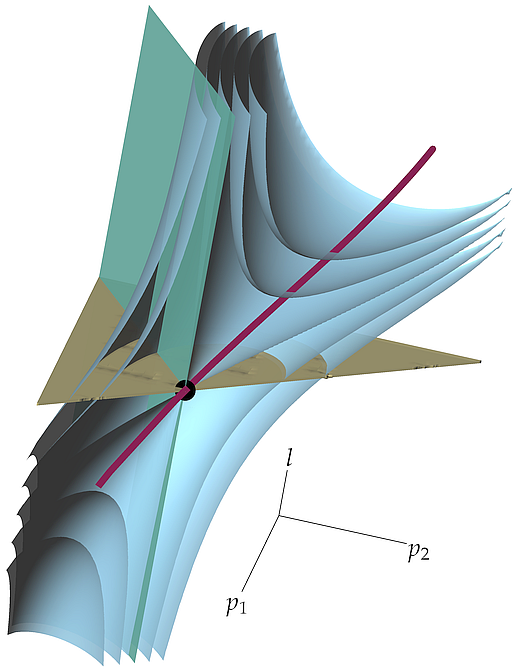}%
\par}\smallskip{\footnotesize
    $K=3$, $N=1$. Equations $\yu\yp-1 =0$ \eqref{eq:con_u} and
    $\ym=1-\ln Z(\yl)$ \eqref{eq:eq_sub_u} are identically satisfied. 
    \textbf{Black dot}: saddle point $(\yp_{\yc},\ym_{\yc},\yl_{\yc})$.\\ 
    \textbf{\textcolor{myblue}{Blue surfaces}}: contours of
    \textcolor{myblue}{$\yF_{\yc}(\yp,\ym,\yl)$} \eqref{eq:Lagrangian}.\\ 
    \textbf{\textcolor{mypurplishred}{Thick purplish-red curve}}: submanifold
    \textcolor{mypurplishred}{$-\ln\yp + \ln\yq + \yl\yM + \ym\yu =0$}
    \eqref{eq:con_p} or
    \textcolor{mypurplishred}{$\yp=\exp(\yl\yM + \ln\yq)/Z(\yl)$}
    \eqref{eq:eq_sub_p}; its intersection with the blue surfaces are the
    \enquote{contours} (points) of $G_{\yc}(\yl)$, \eqn~\eqref{eq:F_from_G}.\\ 
    \textbf{\textcolor{mybluishgreen}{Vertical bluish-green plane}}:
    submanifold
    \textcolor{mybluishgreen}{$\yM\yp-\yc =0$} \eqref{eq:con_l}. \\
    \textbf{\textcolor{mygreenishyellow}{Horizontal greenish-yellow
        plane}}: submanifold
    \textcolor{mygreenishyellow}{$\yM\cd\exp(\yl\yM + \ln\yq)/Z(\yl) =
      \yc$}
    \eqref{eq:eq_sub_l}, which is the plane \textcolor{mygreenishyellow}{$\yl=\yl_{\yc}$}.\par}
\caption{(Colour-blind-friendly palette by Tol \cite*{tol2009_r2012})}
\label{fig:saddle}
\end{figure}

Figure~\ref{fig:saddle} depicts our constrained maximization problem in the
simplest nontrivial case: $K=3$, $N=1$. Our base manifold has therefore $5$
dimensions. The $3$-dimensional space represented in the figure is a
section of our base manifold, obtained by selecting specific values of the
coordinates $p_3$ and $\ym$ as functions -- \eqref{eq:con_p}
and~\eqref{eq:eq_sub_u}--\eqref{eq:Z} below -- of the remaining ones.

The \textbf{\textcolor{myblue}{blue surfaces}} in \fig~\ref{fig:saddle} are
the contour levels of the Lagrangian \eqref{eq:Lagrangian}. Their saddle is
clearly visible. The saddle point is the \textbf{black dot}. Note that a
section roughly parallel to the $\yl$ axis is made within the figure for
clearer visibility of the surfaces and curves involved.

The value of the Lagrangian $\yF_{\yc}$ at the saddle point is the value of
the Gibbs entropy:
\begin{equation}
  \label{eq:G_saddle_is_entropy}
  \yF_{\yc}(\yp_{\yc},\ym_{\yc},\yl_{\yc}) \equiv S(\yc).
\end{equation}
It's easy to check this: the constraint terms in~\eqref{eq:Lagrangian}
vanish at the saddle-point and what remains is the maximized Shannon
entropy, that is, the Gibbs entropy.

\bigskip

Our saddle-point problem reduces, thanks to the continuous
differentiability of the Lagrangian, to the system of three vector
equations
\begin{subequations}\label{eq:con_all}
  \begin{empheq}[left={\mathllap{\begin{aligned}    \de\yF_{\yc}/\de\yp&=0\text{:} \\
        \de\yF_{\yc}/\de\ym&=0\text{:}\\ \de\yF_{\yc}/\de\yl&=0\text{:}\end{aligned}}\qquad}\empheqlbrace]{align}
    \label{eq:con_p}
    -\ln\yp + \ln\yq + \yl\yM + \ym\yu &=0,\\
    \label{eq:con_u}
    \yu\yp-1 &=0,\\
    \label{eq:con_l}
    \yM\yp-\yc &=0.
  \end{empheq}
\end{subequations}
They are the implicit equations of three submanifolds in our
$(K+N+1)$-dimensional base manifold. The first is curved,
$(N+1)$-dimensional. The second is flat, $(K+N)$-dimensional, and contains
the simplex of normalized probability distributions; the $3$-dimensional
space depicted in \fig~\ref{fig:saddle} is the intersection of our
$5$-dimensional base manifold with this submanifold, and therefore
\eqn~\eqref{eq:con_u} is identically satisfied in the figure. The third
submanifold is flat, $(K + 1)$-dimensional; it is the
\textbf{\textcolor{mybluishgreen}{vertical bluish-green plane}} in the
figure. The saddle point is the intersection of these three submanifolds.

The vector equations above can be recombined and written
as a system of three new vector equations:
\begin{subequations}\label{eq:equations_submanifolds}
    \begin{empheq}[left=\empheqlbrace]{gather}
      \label{eq:eq_sub_p}
      \yp=\frac{1}{Z(\yl)}\exp(\yl\yM + \ln\yq),\\
      \label{eq:eq_sub_u}
      \ym=1-\ln Z(\yl),\\
      \label{eq:eq_sub_l}
      \frac{1}{Z(\yl)}\yM\cd\exp(\yl\yM + \ln\yq) = \yc,
    \end{empheq}
    \begin{equation}
      \label{eq:Z}
      \text{with }Z(\yl)\defd \yu\cd\exp(\yl\cd\yM + \ln\yq).
    \end{equation}
\end{subequations}
The first is the parametric equation of a curved $(N+1)$-dimensional
submanifold, which is also submanifold of \eqref{eq:con_u}. The second is
the parametric equation of a curved $(K+N)$-dimensional submanifold; the
$3$-dimensional space of \fig~\ref{fig:saddle} is the intersection of this
submanifold, besides~\eqref{eq:con_u}, with our $5$-dimensional base
manifold; \eqn~\eqref{eq:eq_sub_u} is thus also identically satisfied in
the figure. The third equation in the system above is the implicit equation
of a flat $(K+1)$ dimensional submanifold. This equation determines a
unique value $\yl_{\yc}$, so it is simply the equation of the $(K+1)$-plane
$\yl=\yl_{\yc}$, the \textbf{\textcolor{mygreenishyellow}{horizontal
    greenish-yellow plane}} in \fig~\ref{fig:saddle}. These three
submanifolds \eqref{eq:equations_submanifolds}, are distinct from the
previous three \eqref{eq:con_all}, but they also intersect at the saddle
point. The function $\ln Z(\yl)$ defined in \eqref{eq:Z} is called
\emph{normalization function} or \emph{partition function}.

Note that we could have formulated our constrained-maximization problem by
imposing the normalization~\eqref{eq:con_u} from the very beginning, for
example defining $p_K = 1 - p_1 - p_2 - \dotsb$. The multiplier $\ym$
and \eqns~\eqref{eq:con_u}, \eqref{eq:eq_sub_u} wouldn't have appeared, and
our base manifold would have been $(K+N-1)$-dimensional.
Figure~\ref{fig:saddle} can also be interpreted this way.

\medskip

The system \eqref{eq:con_p} \amp~\eqref{eq:con_u} is equivalent to the
system \eqref{eq:eq_sub_p} \amp~\eqref{eq:eq_sub_u}, as can be verified by
substitution; that is, the $N$-dimensional intersection submanifold of the
first pair is also the intersection of the second pair. This submanifold is
the \textcolor{mypurplishred}{\textbf{thick purplish-red curve}} in the
figure. Projected onto the simplex of probability distributions -- that is,
disregarding the $\ym$ and $\yl$ dimensions -- this submanifold is an
subset of the latter called an \emph{exponential family}. Exponential
families are briefly discussed in \sect~\ref{sec:exponential_families}. The
saddle point is the intersection of this $N$-dimensional submanifold with
either the $(K+1)$-plane \eqref{eq:con_l} or the $(K+1)$-plane
\eqref{eq:eq_sub_l}.

\medskip

The saddle point $(\yp_{\yc}, \ym_{\yc}, \yl_{\yc})$ could therefore be
found by finding the root $\yl=\yl_{\yc}$ of \eqn~\eqref{eq:eq_sub_l}, and
substituting this root in $\bigl( \yp(\yl), \ym(\yl)\bigr)$, the parametric
form of \eqns~\eqref{eq:eq_sub_p} \amp~\eqref{eq:eq_sub_u}. But it turns
out that we do not need to solve \eqn~\eqref{eq:eq_sub_l}. There's an
interesting development.

Looking at \fig~\ref{fig:saddle} we notice that \emph{the $N$-dimensional
  manifold $\bigl( \yp(\yl), \ym(\yl)\bigr)$ extends from the
  pommel 
  to the cantle 
  of the saddle} (if the lower part is the cantle the horse is rearing).
This was a priori not necessary: by construction this submanifold must pass
through the saddle point, but it could have done so by going from the
pommel down to the flaps of the saddle. It couldn't have made a U-turn back to
the pommel, however, because that implies the presence of a wedge,
whereas our manifold is continuously differentiable.

This placement of the $N$-dimensional submanifold implies that if we ride
it we see the values of the Lagrangian decrease until we reach the
saddle point, and then increase again. \emph{The saddle point is therefore
  the minimum of the Lagrangian restricted to the $N$-dimensional
  submanifold.} This is true for the $N=1$ case of our figure, but it
generalizes to larger $N$; here's how.

Consider the restriction of the Lagrangian $\yF_{\yc}$,
\eqn~\eqref{eq:Lagrangian}, to the $N$-dimensional submanifold. This
restriction, denoted $\yG_{\yc}$, is usually called the potential function.
It's by construction a function of $\yl$ alone, from
\eqns~\eqref{eq:eq_sub_p} and \eqref{eq:eq_sub_u}:
\begin{equation}
  \label{eq:F_from_G}
  \yG_{\yc}(\yl) \defd
  \yF_{\yc}[\yp(\yl),\ym(\yl),\yl]
  \equiv \ln Z(\yl) - \yl\cd\yc.
\end{equation}
In \fig~\ref{fig:saddle} the intersections of the blue surfaces with the
thick purplish-red curve are the \enquote{contours} -- just points in this
case -- of this function. The Hessian matrix of its second derivatives has
non-negative eigenvalues: a simple calculation and a look
at~\eqref{eq:equations_submanifolds} reveal that this is in fact the
covariance matrix of the observable $\yM$:
\begin{equation}
  \label{eq:hessian_covariance}
  \frac{\de^2\yG_{\yc}}{\de\yll_n\de\yll_m}
  \equiv \frac{\de^2\ln Z}{\de\yll_n\de\yll_m}
  = \sum_k \ymm^{n k}\ymm^{mk}p_k(\yl)  -
  \sum_k \ymm^{n k}p_k(\yl) \, \sum_k \ymm^{mk}p_k(\yl),
\end{equation}
and covariance matrices have non-negative eigenvalues
\citep[\sect~III.5]{feller1966_r1971}. The potential function $\yG_{\yc}$
is therefore \emph{convex} -- strictly so, without flat regions, owing to the
differential properties of the logarithm. Calculation of its unique minimum
by derivation leads to \eqn~\eqref{eq:eq_sub_l}. The conclusion is that
\emph{the potential function $\yG_{\yc}$ is convex in $\yl$, and the saddle
  point of $\yF_{\yc}$ in the $(K+N+1)$-dimensional base manifold is the
  unique minimum of $\yG_{\yc}$ in the $N$-submanifold}.

This is the geometric reason why the constrained-maximization problem in
$K$ dimensions for the Shannon entropy $H(\yp)$ can be transformed into an
unconstrained-minimization problem in $N$ dimensions for the potential
function $\ln Z(\yl) - \yl\cd\yc$. The latter is usually called the
\emph{dual problem} \citep{fangetal1997,boydetal2004_r2009}. This fact is
enormously useful for numerical computations: it allows us to use convex
optimization techniques \citep{pressetal1988_r2007} to find the extremizing
Lagrange multipliers $\yl_{\yc}$ and thence the distribution $\yp_{\yc}$
and the Gibbs entropy $S(\yc) \defd H(\yp_{\yc})$.
See Rockafellar's \citey{rockafellar1993} insightful discussion in this respect too.

But the geometry of this extremization problem has further surprises.

\medskip

The Gibbs entropy $S(\yc)$ is, from its definition, equal to
$\yF_{\yc}(\yp_{\yc},\ym_{\yc},\yl_{\yc}) \equiv \yG_{\yc}(\yl_{\yc})
\equiv \ln Z(\yl_{\yc}) - \yl_{\yc}\cd\yc$, which is the unique minimum of
$\yG_{\yc}(\yl)$. We can write this as
\begin{equation}
  \label{eq:F_lagrange_S}
  S(\yc) 
  = \inf_{\smash{\yl}}[\ln Z(\yl) - \yl\cd\yc].
\end{equation}
This formula is the proper definition of the negative \emph{Legendre
  transform} of the normalization function $\ln Z(\yl)$, that is, its
negative convex conjugate \citep{fenchel1949}. This means that \emph{the
  Gibbs entropy $S(\yc)$ is a concave function of $\yc$}. The concavity of
the Gibbs entropy is an important property, completely distinct from the
concavity of the Shannon entropy $H(\yp)$. I said \enquote{proper} Legendre
transform because the physics literature often defines the latter without
the extremization indicated by \enquote{$\inf$} or \enquote{$\sup$}. Such
simplified definition breaks down if the transformed function is
not strictly convex -- which may happen in important physical situations.
See Wightman's illuminating and pedagogical discussion
\citep[pp.~xxiv--xxix]{wightman1979}.

From the involutive property of the Legendre transform \parentext{see refs
  above}, \emph{the normalization function $\ln Z(\yl)$ is the negative Legendre
  transform of the negative Gibbs entropy}:
\begin{equation}
  \label{eq:F_lagrange_Z}
  \ln Z(\yl) = \inf_{\smash{\yc}}[-S(\yc) - \yc\yl].
\end{equation}

The appearance of these Lagrange transforms has many important connections
with statistical mechanics; for interesting recent developments see Chomaz
\amp\ al.'s \citey{chomazetal2006,chomazetal2005b} reviews. I refrain from
speaking about the relationship between the maximum-entropy method: it's
subtle and already too often oversimplified in the literature.

\section{Synopsis of the main functions} It's important to keep the main
functions involved in the proof well-distinct from one another:

\begin{itemize}[para]
\item  The \emph{Shannon entropy} $H(\yp)$ is a function of the probability
  distribution $\yp$. It's concave in $\yp$. It isn't the Legendre
  transform of anything.

\item  The \emph{Gibbs entropy}
  \begin{equation*}
    S(\yc) \equiv
    \sup_{\smash{\yp}}^{\smash{\yM\yp=\yc}}H(\yp)
    \equiv \inf_{\smash{\yl}}[\ln\yu\cd\exp(\yl\cd\yM + \ln\yq)- \yl\yc]
    \equiv  \inf_{\smash{\yl}}[\ln Z(\yl)- \yl\yc]
  \end{equation*}
  is a function of the expectation values $\yc$. It's concave in $\yc$.
  It's the constrained maximum of the relative Shannon entropy, the
  unconstrained minimum of the potential function, and the negative
  Legendre transform of the normalization function.

\item  The \emph{normalization function}, \emph{partition function}, or
  \emph{free entropy}
  \begin{equation*}
    \ln Z(\yl) \equiv
    \ln\yu\cd\exp(\yl\cd\yM + \ln\yq) \equiv
    \inf_{\smash{\yc}}[-S(\yc) - \yc\yl]
  \end{equation*}
  is a function of the Lagrange multipliers $\yl$. It's convex in $\yl$.
  It's the negative Legendre transform of the negative Gibbs entropy.

\item  The \emph{potential function} $\yG_{\yc}(\yl) \equiv \ln Z(\yl)- \yl\yc$
  is a function of the Lagrange multipliers $\yl$ with a parametric
  dependence on the expectation values $\yc$. It's convex in $\yl$. It
  isn't the Legendre transform of anything. Not to be confused with the
  Gibbs entropy.

\end{itemize}


\section{Exponential families}\label{sec:exponential_families}

The maximum-entropy method is essentially a function that maps a set of
observables, a set of observable constraints, and a reference
distribution to a probability distribution: $(\yM,\yc,\yq) \mapsto \yp$.
From this point of view all other quantities appearing in its proof and
formulae are just auxiliary quantities -- including the Lagrange
multipliers $\yl$.

But the parametric submanifold of probabilities $\yp(\yl)$,
\eqn~\eqref{eq:eq_sub_p}:
\begin{equation}
  \label{eq:exponential_family}
 \yp(\yl)=\frac{1}{Z(\yl)}\exp(\yl\yM + \ln\yq), \qquad
      Z(\yl)\defd \yu\cd\exp(\yl\cd\yM + \ln\yq),
\end{equation}
has a meaning and an importance of its own, outside of the maximum-entropy
method. It is an example of \emph{exponential family}. Exponential families
are particular submanifolds of a simplex of probability distributions
characterized by an exponential parametric form like the above or more
general \parentext{see below for references}.

\setlength{\intextsep}{0.5ex}%
  \begin{wrapfigure}{r}{0.5\linewidth}
  \centering\includegraphics[width=\linewidth]{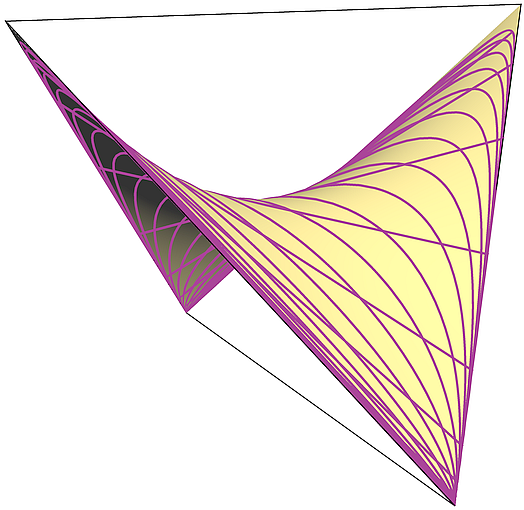}\\
  \includegraphics[width=\linewidth]{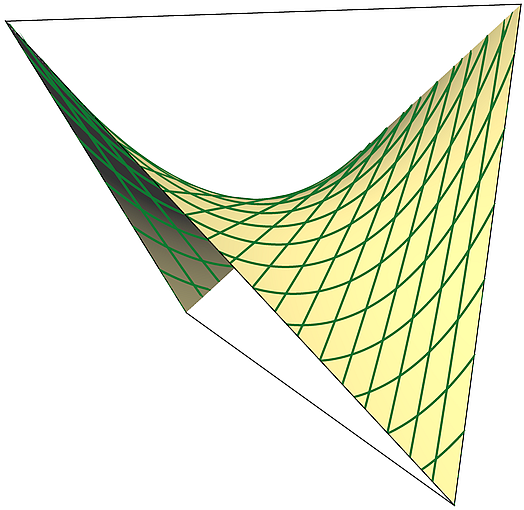}
\end{wrapfigure}
The figures on the right show an example of $2$-dimensional exponential
family for the case with four states, $K=4$, two observables, $N=2$, having values
\begin{equation}
  \label{eq:observables_expfamily_example}
  (\ymm^{nk}) \equiv
  \begin{pmatrix}
    \ymm^{1\,k}\\\ymm^{2\,k}
  \end{pmatrix}
=
  \begin{pmatrix}
    0 & 1 & 2 & 3\\ 1& 1& 0 & 0
  \end{pmatrix},
\end{equation}
and a uniform reference distribution $\yq$. The \textbf{black lines} are the
edges of the \textbf{simplex of probability distributions} $\yp$, a
tetrahedron. The \textbf{\textcolor{myyellow}{exponential-family
    submanifold}} is the \textbf{\textcolor{myyellow}{yellow surface}}, the
same in both figures.

In the top figure the \textbf{\textcolor{myreddishpurple}{reddish-purple
    curves}} are equally-spaced
\textbf{\textcolor{myreddishpurple}{$(\yll_1, \yll_2)$ coordinates}} for
the parametrization in terms of the Lagrange multipliers,
$\yl\defd(\yll_1, \yll_2) \in \RR^2$. In the bottom figure the
\textbf{\textcolor{mygreen}{green curves}} are equally-spaced
\textbf{\textcolor{mygreen}{$(\ycc_1, \ycc_2)$ coordinates}} for the
parametrization in terms of the expectations,
$\yc \defd (\ycc_1, \ycc_2) \in \clcl{0,3}\times\clcl{0,1}$.

The relation between these two coordinate systems, \eqn~\eqref{eq:con_l},
is highly non-linear. The $\yl$ coordinates have the advantage of
parametrizing $\yp(\yl)$ in closed form, \eqn~\eqref{eq:eq_sub_p}, but are
uncongenial to the convex structure of the simplex of probability
distributions; their non-compact range must in fact cover a compact set.
The $\yc$ coordinates are clearly more congenial to the convex structure of
the probability simplex, but they do not lend themselves to a
parametrization $\yp(\yc)$ in closed form. Both $\yl$- and
$\yc$-parametrizations are therefore important.

The Bernoulli, Poisson, exponential, normal distributions belong to
exponential families. See Bernardo \amp\ Smith \citey[\chap~4, esp.\
\sect~4.5.3]{bernardoetal1994_r2000} for a thorough discussion of these
families, Barndorff-Nielsen \citey{barndorffnielsen1978_r2014} for their
relation with Lagrange transforms, Dawid \citey{dawid2013} for a broader
context. Exponential families appear in the probability calculus when we
assume that a particular fixed set of quantities from some measurements is
all we need to make inferences about other similar measurements, a
condition called \emph{sufficiency} \cites[\chap~4, esp.\
\sect~4.5.3]{bernardoetal1994_r2000}{barndorffnielsen1978_r2014,dawid2013,andersen1970}.
The fact that they appear in the maximum-entropy formulae thus suggests a
relation between maximum-entropy and sufficiency. A recent work which I
don't fully understand \citep{portamana2017} argues, however, that this
relation has some downsides, and maximum-entropy distributions are best
related to so-called \emph{exchangeable} models
\citep[\sect~4.3]{bernardoetal1994_r2000}.



\ifnotnotes
\begin{acknowledgements}
  I owe the inspiration for writing this note to Moritz Helias, Tobias
  K{\"u}hn, and Vahid Rostami. I cordially thank Tobias for detecting some
  deficiencies in an early draft. It goes without saying that any 
  deficiencies that may remain are therefore \emph{his} fault, right?
  \ldots ah,
  wait, it doesn't work that way?
\end{acknowledgements}
\fi



\defbibnote{prenote}{{\footnotesize (\enquote{van $X$} is listed under V;
    similarly for other prefixes, regardless of national
    conventions.)\par}}

\printbibliography[
]

\end{document}